\documentclass[10pt,twocolumn,letterpaper]{article}

\usepackage[pagenumbers]{cvpr} 

\usepackage{multirow}     
\usepackage{booktabs}     
\usepackage{array}        
\usepackage{pifont}       
\usepackage{colortbl}     
\usepackage{fontawesome5}

\usepackage[most]{tcolorbox} 
\usepackage{enumitem}        
\usepackage{listings}        
\usepackage{xspace}          

\providecommand{\etc}{\textit{etc}.\xspace}

\definecolor{jsonbg}{rgb}{0.97,0.97,0.97}
\lstdefinestyle{jsonstyle}{
    basicstyle=\ttfamily\small,
    backgroundcolor=\color{jsonbg},
    breaklines=true,
    showstringspaces=false,
    columns=fullflexible,
    frame=none,
    xleftmargin=2pt,
    xrightmargin=2pt,
}

\definecolor{cvprblue}{rgb}{0.21,0.49,0.74}
\usepackage[pagebackref,breaklinks,colorlinks,allcolors=cvprblue]{hyperref}

\def\paperID{*****} 
\def\confName{CVPR}
\def\confYear{2027}

\title{MAPS: A Multi-Agent Personalized Script-Driven Video Advertising System Enhanced by Reinforcement Learning}

\author{
  \textbf{Zhongyu Wang}\textsuperscript{1,2,*}
  \quad \textbf{Chen Zuo}\textsuperscript{2,*}
  \quad \textbf{Qingxuan Sun}\textsuperscript{2,*}
  \quad \textbf{Jiayang Qiao}\textsuperscript{2} \\
  \textbf{Qiuhao Qian}\textsuperscript{2,\dag}
  \quad \textbf{Songyan Cui}\textsuperscript{2}
  \quad \textbf{Guocheng Chen}\textsuperscript{2} \\[6pt]
  \textsuperscript{1}Beihang University \quad \textsuperscript{2}Kuaishou Technology
}

\begin{document}
\maketitle
\begingroup
\renewcommand\thefootnote{}\footnotetext{
\hspace*{-1.8mm}\begin{tabular}{@{}l@{}}
$^{*}$Equal contribution.\\
$^{\dagger}$Corresponding author.
\end{tabular}
}
\endgroup

\begin{abstract}

With the rapid growth of online video consumption, video advertising has become a dominant component of digital advertising. Most personalized video advertising systems follow a retrieval-based paradigm, which matches ad content with user preferences in a feature space and recommends the most similar candidates. However, since the candidate creatives are pre-produced, this paradigm limits the granularity of personalization and leads to performance disparities across diverse user groups. To address this limitation, we propose a multi-agent personalized script-driven video advertising system (MAPS). We design four collaborative agents that form a closed-loop pipeline of user profile modeling, script generation, video synthesis, and feedback-driven iterative refinement. To train the Scripter Agent, we develop a reinforcement learning framework equipped with a multi-dimensional reward function, which steers the agent toward producing scripts that simultaneously align with user content preferences and target product information. Extensive offline experiments on representative datasets demonstrate that our approach achieves encouraging performance in generating and integrating personalized creatives. Online A/B experiments in a specific scenario in the Kuaishou advertising system show that MAPS improves CTR by 2.87\%, CVR by 2.74\%, and ARPU by 9.85\% over the generative baseline.
\end{abstract}    
\section{Introduction}
\label{sec:introduction}

\begin{figure}[t]
  \centering
  \includegraphics[width=0.95\columnwidth]{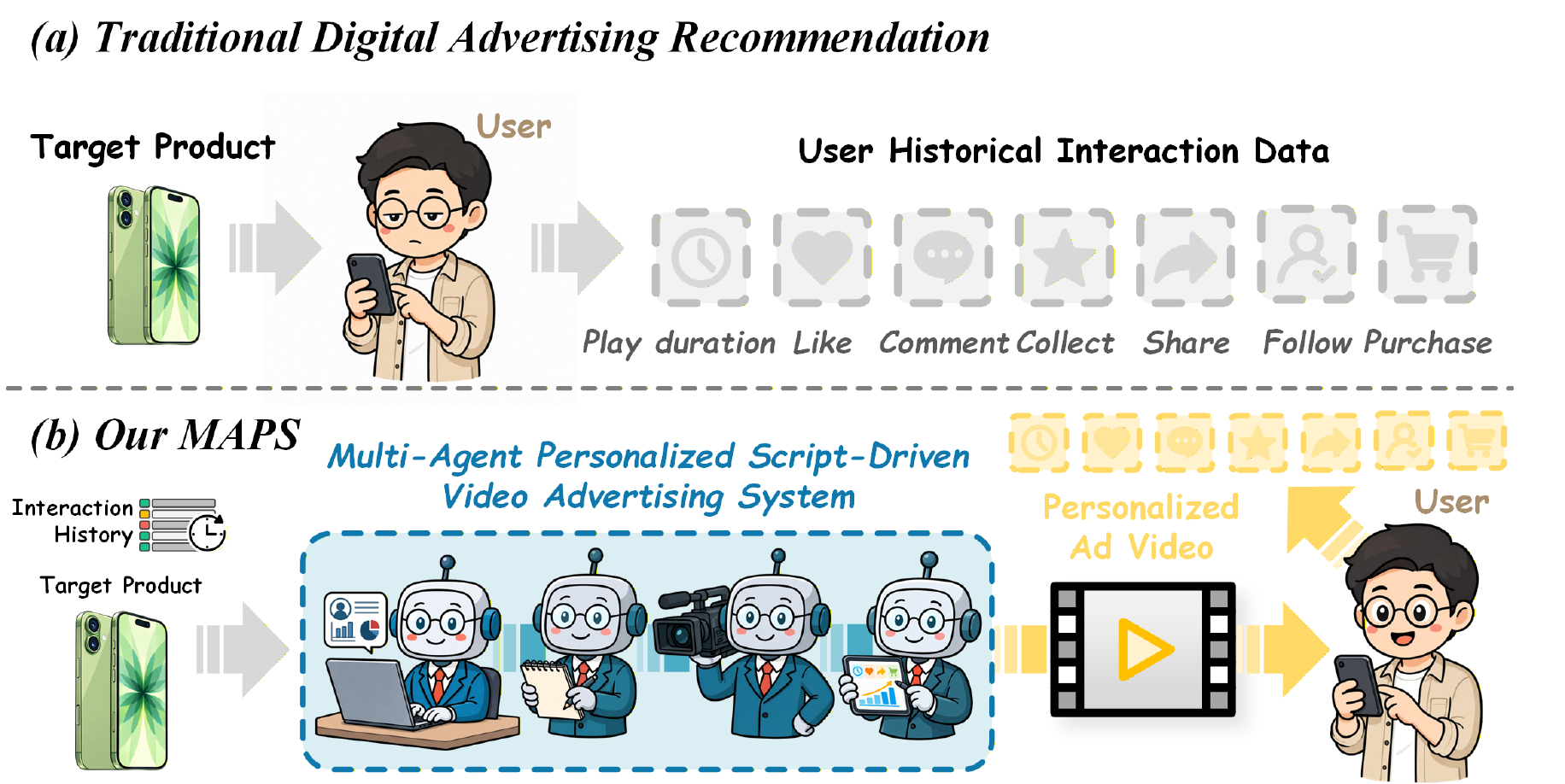}
  \caption{Comparison between our MAPS and existing digital advertising recommendation systems.}
  \label{fig:teaser}
\end{figure}

{\noindent\itshape
``Once you crack the script, everything else follows.''%
\begin{flushright}
\normalfont--- Ridley Scott
\end{flushright}
}

As the financial backbone of the modern internet ecosystem, digital advertising serves as the primary revenue engine for social media and content platforms~\cite{Ji_2026_CVPR, peng2026r, li2025optimal}. Compared to text-based and static image advertisements, video enables richer narrative expression and an enhanced sense of immersion, fostering deeper emotional resonance and stronger user interaction~\cite{Zhou_2026_CVPR, xu2026ad, song2025addriftbench}. Consequently, video advertising has emerged as one of the fastest-growing segments within the digital advertising landscape. Over the past several decades, advertising optimization research has been primarily concentrated on targeting, allocation, and bidding mechanisms~\cite{yang2026ahbid, liaw2025risk, li2025beyond}. As these techniques approach a performance plateau, the next frontier of efficacy improvement has shifted toward the creative itself. Creative optimization grounded in user characteristics and contextual information is increasingly recognized as the pivotal determinant of advertising effectiveness~\cite{wang2026mdocrag, Xu_2026_CVPR, zhang2026beyond, zhang2025efficient}.

Nevertheless, video creatives are costly artifacts produced by specialized teams, characterized by limited output and extended production cycles, which collectively constrain both the quantity and diversity of deployable creatives~\cite{ma2025genauction, feng2025understanding}. Under these constraints, the industry has converged on a retrieval-based paradigm: advertisers typically commission professional personnel to pre-produce a limited pool of candidate video creatives or modular assets targeting specific products; at serving time, the most suitable variant is retrieved to match user preferences and contextual information~\cite{zheng2026chasm, morio2026multimodal, soumalias2026truthful}. Despite its practical utility, this paradigm suffers from two fundamental limitations. First, static inventory constraints: the static and finite nature of the creative repository renders it inadequate to accommodate rapidly evolving user preferences, thereby compromising user resonance and overall advertising effectiveness. Second, feedback latency: user feedback signals can only inform subsequent production cycles and offer no mechanism for real-time optimization of creatives currently in deployment, introducing a significant latency between performance signals and creative production.

To address these limitations, we propose MAPS, a multi-agent personalized script-driven video advertising system. MAPS consists of four collaborative agents. The Profile Agent constructs structured user profiles from historical interaction data and contextual information, including the video currently being consumed by the user, and continuously refines these profiles as new behavioral observations accumulate. The Scripter Agent generates personalized advertisement video scripts conditioned on user preferences, guided by target product information such as core selling points and available creative assets. The Director Agent synthesizes personalized candidate creatives from the generated scripts and target product information, ensuring that the outputs simultaneously align with individual user preferences and faithfully convey target product information. The Reflector Agent attributes online user interaction signals to profile-level and script-level causes and converts them into minimal revisions routed back to the Profile and Scripter Agents for iterative refinement. To train the Scripter Agent, we construct a reinforcement learning framework based on the Group Relative Policy Optimization (GRPO)~\cite{shao2024deepseekmath} algorithm and design a multi-dimensional reward function that ensures structural output regularity while steering the generated scripts to align advertisement narratives with target product information and individual user preferences. Offline experiments on two representative tasks demonstrate that MAPS captures user preferences with high fidelity and generates high-quality, personalized, and native video creatives. Online A/B experiments in a specific scenario in the Kuaishou advertising system further demonstrate that MAPS delivers significant improvements in click-through rate (CTR), conversion rate (CVR), and average revenue per user (ARPU) over the generative baseline at industrial scale.

Our main contributions are summarized as follows:

\begin{itemize}
    \item We propose \textsc{MAPS}, a multi-agent personalized script-driven video advertising system that fundamentally transforms personalized video advertising from static retrieval over a discrete creative inventory into dynamic generative optimization across a continuous creative space.

    \item We establish a closed-loop creative evolution mechanism through post-deployment online user feedback, enabling advertisement creatives to be iteratively refined across successive impressions in response to real-time user interactions.

    \item We develop a reinforcement learning framework for the Scripter Agent, incorporating a multi-dimensional reward function that jointly steers the generated scripts to align with individual user preferences and faithfully reflect target product information.

    \item We demonstrate the effectiveness of MAPS through extensive offline experiments, and further validate its practical value at industrial scale through online A/B experiments in a specific scenario in the Kuaishou advertising system.
\end{itemize}

\section{Related Work}
\label{sec:related_work}

\subsection{Video Advertising}
The evolution of video advertising can be characterized along two critical dimensions: personalization and nativeness~\cite{long2025adsqa, qian2025vc, aggarwal2024auto}. Traditional advertising formats, such as billboards and broadcast television commercials, occupy the lower end of both dimensions, relying on static creatives and mass-reach distribution. Digital advertising, encompassing programmatic display, search, and retargeting, primarily advances personalization through audience targeting, yet typically achieves only limited integration between creatives and contextual content. Native advertising formats, including product placement, branded content marketing, and influencer marketing, attain a high degree of nativeness through narrative expression and seamless contextual integration, but remain constrained in their capacity for individualized personalization. Scaling creative-level personalization has proven particularly challenging due to prohibitive production costs and labor-intensive creative workflows~\cite{aggarwal2026online, pelissier2026users, el2026contextual}. Although dynamic creative optimization partially alleviates this tension, the substantial overhead of video production continues to impede the simultaneous realization of high personalization and high nativeness. To address this challenge, we propose MAPS, a multi-agent system that transforms personalized video advertising from static retrieval over a discrete creative inventory into dynamic generative optimization across a continuous creative space, simultaneously achieving high personalization and high nativeness within a unified framework.

\begin{figure*}[t]
    \centering
    \includegraphics[width=\textwidth]{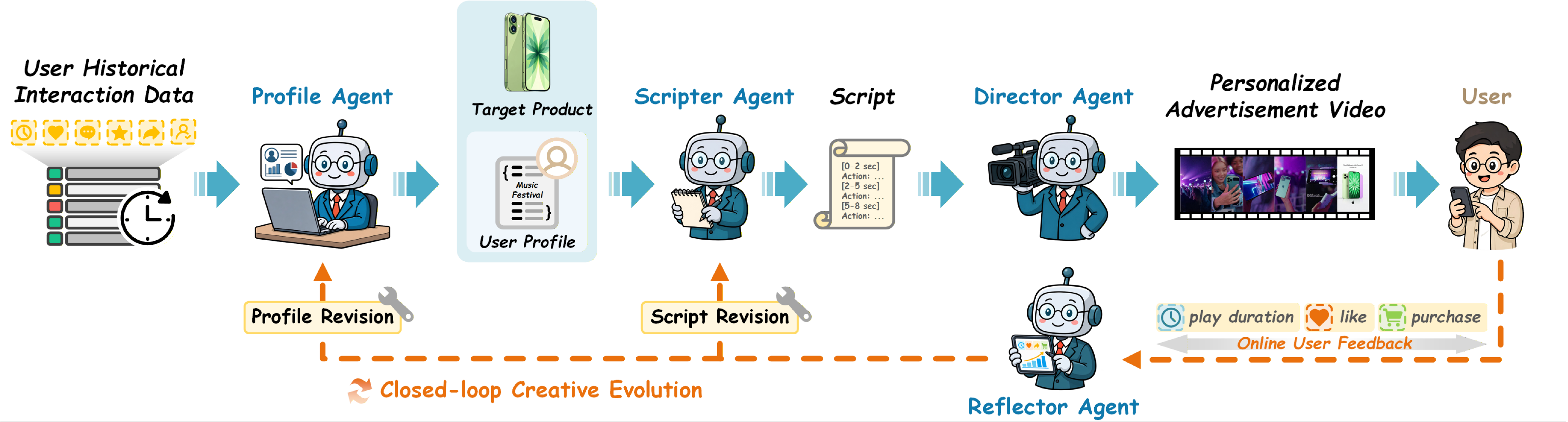}
    \caption{Overview of the MAPS framework.}
    \label{fig:maps_framework}
\end{figure*}

\subsection{Multi-Agent Systems in Personalized Video Advertising}
Recent advances in generative artificial intelligence have made large-scale synthesis of advertisement creatives increasingly viable~\cite{chen2026refadgen, huang2025generative, hou2025leveraging, zhang2026symphomotion, xiang2025macro}. Prior work has leveraged diffusion models and large language models (LLMs) to generate personalized advertisement images and banner creatives, and has incorporated interaction-oriented objectives, such as click-through rate, to guide the generation process toward utility-driven optimization~\cite{tam2026vietcheckmed, song2026dara, tessr, chenmathcal, kimpre, li2026towards, liu2025real, wang2025icl}. Multi-agent systems built upon multimodal large language models (MLLMs) have demonstrated promise in personalized advertising workflows, benefiting from the collaboration among specialized components~\cite{wang2026mars, wang2026gramrag, wang2025banneragency, kruger2025improving, shi2025personax}. However, existing works remain predominantly centered on static image creatives, while generative approaches to personalized video advertising have received less attention. To bridge this gap, MAPS introduces a script-driven, closed-loop multi-agent framework that positions the script as the central interface between planning and generation, and trains the Scripter Agent via reinforcement learning to jointly optimize structural regularity, user preference conformity, and product alignment.
\section{Method}
\label{sec:method}

\subsection{Architecture Overview}

As illustrated in Figure~\ref{fig:maps_framework}, MAPS consists of four collaborative agents: a Profile Agent that extracts user preferences, a Scripter Agent that generates structured scripts, a Director Agent that produces personalized advertisement videos from the generated scripts, and a Reflector Agent that attributes online user feedback to profile-level and script-level causes and routes iterative refinements back to the Profile and Scripter Agents.

\begin{figure*}[t]
    \centering
    \includegraphics[width=\textwidth]{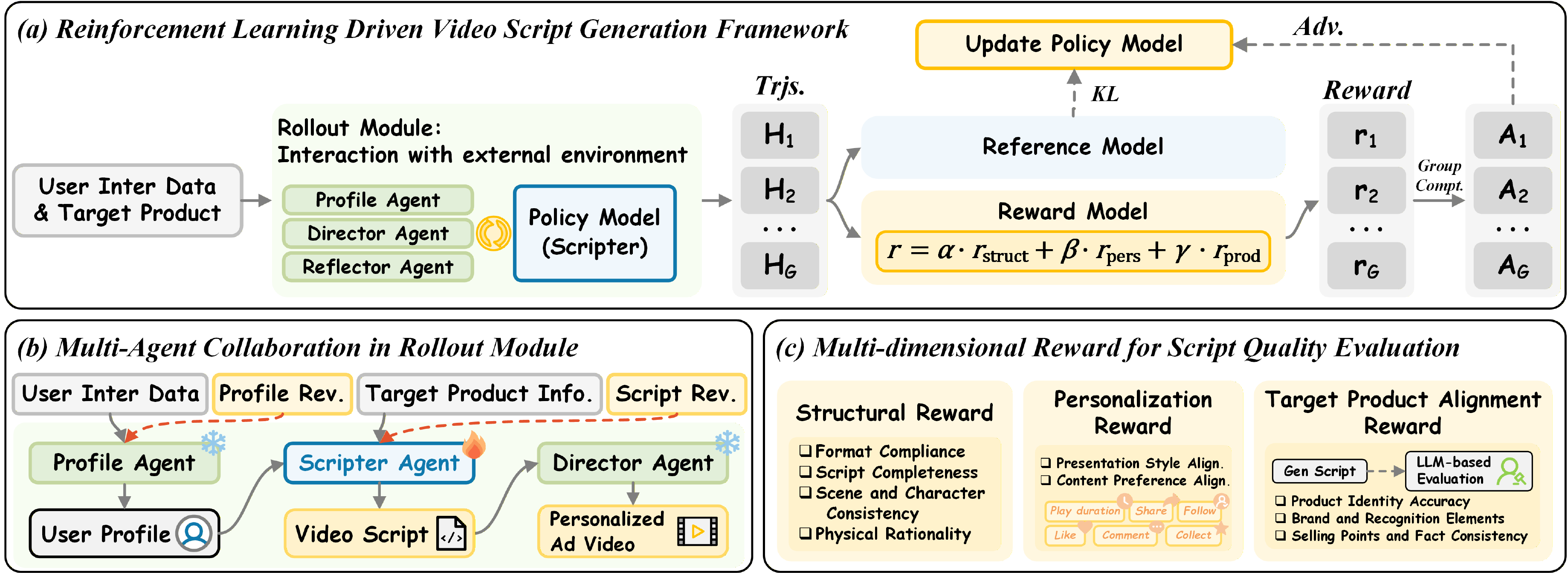}
    \caption{Overview of our reinforcement learning framework.}
    \label{fig:grpo}
\end{figure*}

\subsection{Profile Agent}

The Profile Agent serves as the perceptual and modeling foundation of the MAPS framework. Its primary responsibility is to construct structured user profiles from heterogeneous historical interaction data and real-time contextual signals, and to continuously refine these profiles as new behavioral observations accumulate, thereby providing high-quality user preference priors for the Scripter Agent's subsequent personalized script generation.

\subsubsection{Input.}
The inputs to the Profile Agent encompass two categories of information.
The first category comprises user historical interaction data, including viewing history, play duration, likes, comments, collections, shares, and follows.
These multi-dimensional behavioral signals collectively characterize the user's historical content consumption patterns.
The second category consists of real-time contextual information, specifically the video content currently being consumed by the user, which serves as an immediate contextual anchor for preference inference.
Both categories of information are jointly fed into the Profile Agent, which employs an MLLM as its backbone to drive the profile construction process.

\subsubsection{User Profile Construction.}
Based on the above inputs, the Profile Agent performs deep inference over user content consumption preferences and produces a structured user profile across four dimensions:

\begin{itemize}
    \item \textbf{Topical interests}: the content domains and themes preferred by the user, such as gaming, travel, music, and fashion;

    \item \textbf{Preferred presentation styles}: the content expression styles favored by the user, such as humorous, informative, emotionally resonant, or narratively immersive;

    \item \textbf{Implicit needs}: latent needs revealed through the user's content consumption behaviors, such as quality-of-life improvement, self-expression, and entertainment;

    \item \textbf{Decision-making tendencies}: behavioral preferences exhibited in consumption decisions, such as price-sensitivity, quality-orientation, or trend-following.
\end{itemize}

The resulting multi-dimensional profile is serialized into structured text and passed to the Scripter Agent as the user preference conditioning for script generation.

\subsubsection{Dynamic Refinement.}
The user profile is not a static artifact but evolves as behavioral observations continue to accumulate. 
Following each advertisement impression, the Reflector Agent attributes online user feedback to profile-level causes and routes the resulting revisions to the Profile Agent, which incorporates them into its update procedure to drive iterative refinement of the user profile. This dynamic update mechanism enables the system to responsively capture shifts in user preferences, thereby supporting closed-loop creative evolution across successive impressions over extended temporal horizons and facilitating progressive narrative development.

\subsection{Scripter Agent}

The Scripter Agent is the core creative generation module of the MAPS framework, responsible for transforming target product information into a structurally complete, personalized advertisement video script conditioned on individual user characteristics.
It receives two streams of input: the structured user profile output by the Profile Agent, which encapsulates the user's interest preferences, aesthetic inclinations, and consumption motivations; and the detailed attribute information of the target product, including product appearance, core selling points, and brand recognition elements.
Conditioned on these two input streams, the Scripter Agent produces a fine-grained personalized advertisement script that fully specifies the script content, voiceover, on-screen text, visual descriptions, and pacing for each scene, and passes the resulting script to the Director Agent for subsequent video production. 
In addition, the Scripter Agent receives script-level revisions routed from the Reflector Agent and incorporates them to iteratively refine the script.

Although general-purpose MLLMs possess a certain baseline capability for script generation, crafting scripts for personalized short-form advertisement videos demands more specialized competencies: naturally weaving product information into narrative contexts aligned with user preferences, adapting tone and expressive style to individual user characteristics, and constructing emotionally resonant call-to-action (CTA) sequences that match user decision-making tendencies.
To acquire these competencies, we adopt a two-stage training paradigm: a supervised fine-tuning (SFT) stage to establish the foundational capability for structured script generation, followed by a reinforcement learning (RL) stage to align model outputs with personalized quality objectives.

\subsubsection{Stage One: Supervised Fine-Tuning for Structural Capability.}
The SFT stage aims to teach the model the fundamental syntax and structural conventions of advertisement scripts within the MAPS framework. We formulate the task as a sequence-to-sequence generation problem: the input $x$ consists of the structured user profile and target product attribute information, while the output $y$ is the structured text of the target advertisement script. We fine-tune a base language model on a corpus $\mathcal{D}$ of human-authored personalized advertisement scripts, with the training objective of maximizing the conditional log-likelihood of the reference scripts:
\begin{equation}
    \mathcal{L}_{\text{SFT}}
    = -\mathbb{E}_{(x,y) \sim \mathcal{D}}
    \left[
        \sum_{t=1}^{|y|} \log \pi_{\theta}(y_t \mid y_{<t},\, x)
    \right]
\end{equation}
The SFT cold start stage equips the model with the foundational ability to generate structurally complete scripts that are consistent with the user profile and product attribute information, providing a solid basis for the subsequent personalization alignment stage.

\subsubsection{Stage Two: Reinforcement Learning for Personalization Alignment}
While SFT cold start guarantees structural completeness and content correctness of the generated scripts, it falls short of capturing the fine-grained alignment quality required for personalized short-form advertisement creation, namely the tight correspondence between script content and the core information of the target product, as well as the deep alignment between narrative style and individual user preferences. To address this, we introduce a reinforcement learning framework using the GRPO algorithm as illustrated in Figure~\ref{fig:grpo}.

The RL stage employs a hybrid reward function $R_{\text{total}}$ composed of three complementary components:

\begin{itemize}

\item \textbf{Rule-Based Structural Reward ($R_{\text{struct}}$)}: This component provides an objective signal for the technical correctness of the generated script, evaluating and aggregating normalized scores across four automated checks: Format Compliance (whether the script conforms to the required structural format), Script Completeness (whether all required script components are fully present and accounted for), Scene and Character Consistency (whether scene arrangement and character settings remain coherent throughout), and Physical Rationality (whether character positions and camera geometry are plausible within real-world constraints).

\item \textbf{Personalization Reward ($R_{\text{pers}}$)}: This component measures the degree to which the generated script resonates with the user profile, assessed via an LLM-as-a-Judge paradigm across two core dimensions: presentation style alignment and content preference alignment.
For presentation style alignment, the LLM judge evaluates the script against the user's inferred aesthetic preferences derived from their behavioral history, considering four aspects: tone (whether it matches the user's preferred warm, cold, or neutral style), lighting treatment (whether it suits the user's preference for soft, harsh, or natural lighting), textural style (whether it fits the user's aesthetic inclination), and artistic style (whether it aligns with the user's taste for minimalistic, modern, vintage, or other aesthetics).
For content preference alignment, the LLM judge assesses subject matter relevance (whether the script theme matches the user's interest domains), scene semantics (whether scene selection fits the user's preferences), information angle (whether the product messaging focuses on aspects the user cares about, such as performance, aesthetics, or value), and motivational appeal (whether the ad's appeal resonates with the user's driving motivations such as quality, efficiency, or lifestyle).
The LLM judge outputs a score on a scale of 0 to 10 for each dimension, which are normalized and aggregated to form $R_{\text{pers}}$.

\item \textbf{Product Alignment Reward ($R_{\text{prod}}$)}: This component measures the consistency between the product depicted in the generated script and the actual attributes of the target product, also assessed via an LLM-as-a-Judge paradigm.
The LLM judge is provided with detailed product information alongside the generated script and evaluates three dimensions: product identity accuracy (whether the script correctly portrays the product's appearance, color, model, packaging, and accessories), brand and recognition elements (whether the script accurately represents the product's logo, brand visual language, and design style), and selling points and factual consistency (whether the product claims conveyed in the script accurately reflect the real product features without exaggeration or misrepresentation).
The LLM judge outputs a score on a scale of 0 to 10, which is normalized to serve as the $R_{\text{prod}}$ signal.

\end{itemize}
The final reward function integrates all three components through a weighted combination:
\begin{equation}
    R_{\text{total}}(y) = \alpha \cdot R_{\text{struct}}(y)
    + \beta \cdot R_{\text{pers}}(y)
    + \gamma \cdot R_{\text{prod}}(y)
\end{equation}
where $\alpha = \text{0.3}$, $\beta = \text{0.4}$, and $\gamma = \text{0.3}$. During optimization, for each input $x$, $K$ candidate scripts $\{y^{(k)}\}_{k=1}^{K}$ are sampled from the current policy $\pi_\theta(\cdot \mid x)$, and their rewards $R_k = R_{\text{total}}(y^{(k)})$ are computed.
A normalized advantage is then calculated within the group:
\begin{equation}
    A_k = \frac{R_k - \bar{R}}{\sigma_R + \varepsilon},
    \qquad
    \bar{R} = \frac{1}{K}\sum_{k=1}^{K} R_k
\end{equation}
The policy is updated by maximizing the advantage-weighted log-likelihood, constrained by a KL-divergence penalty to prevent large deviations from the SFT initialization:
\begin{equation}
\begin{split}
    \mathcal{L}_{\text{GRPO}} &=
    \mathbb{E}_{x\sim\mathcal{D}} \!\left[
        \frac{1}{K}\sum_{k=1}^{K} A_k \cdot \log \pi_\theta\!\left(y^{(k)} \mid x\right)
    \right] \\
    &- \beta_{\text{KL}} \cdot \mathbb{E}_{x\sim\mathcal{D}} \!\left[
        \mathrm{KL}\!\left(
            \pi_\theta(\cdot \mid x) \,\|\, \pi_{\text{SFT}}(\cdot \mid x)
        \right)
    \right]
\end{split}
\end{equation}
Through the above two-stage training paradigm, the capability of the Scripter Agent advances from generating structurally well-formed advertisement scripts to producing compelling personalized advertisement creatives that align tightly with the core information of the target product and match deeply with individual user preferences.

\begin{table*}[t]
    \small
    \centering
    \resizebox{1.00\textwidth}{!}{
    \begin{tabular}{l|cc|c|c|c|cccc}
    \toprule
    \multirow{2}{*}{\textbf{Method}} & 
    \multicolumn{2}{c|}{\textbf{Personalization}} &
    \multicolumn{1}{c|}{\multirow{2}{*}{\textbf{Product Alignment}}} &
    \multicolumn{1}{c|}{\multirow{2}{*}{\textbf{Integration Quality}}} &
    \multicolumn{1}{c|}{\multirow{2}{*}{\textbf{Diversity}}} &
    \multicolumn{4}{c}{\textbf{Video Quality}} \\
    & \textbf{Presentation} & \textbf{Content} &  &  &  & \textbf{MS} & \textbf{DD} & \textbf{AQ} & \textbf{IQ} \\
    \midrule
    \multicolumn{10}{c}{$\textit{PCG-Bench}$}\\
    \midrule
    GenericAds (Closed) & 1.78 & 1.55 & 8.62 & -- & 7.15 & 0.9737 & 0.9470 & 0.5427 & 0.6564  \\
    NextAds (Closed) & 7.17 & 5.21 & 8.68 & -- & 6.69 & 0.9473 & 0.9520 & 0.5301 & 0.6287  \\
    NextAds (Open) & 6.00 & 3.39 & 3.53 & -- & 5.52 & 0.9482 & 0.9526 & 0.4025 & 0.6364  \\
    \rowcolor{gray!15} \textbf{MAPS} & \textbf{7.32} & \textbf{5.35} & \textbf{8.89} & {--} & \textbf{7.30} & \textbf{0.9862} & \textbf{0.9871} & \textbf{0.5603} & \textbf{0.6690}  \\
    \midrule
    \multicolumn{10}{c}{$\textit{PCI-Bench}$}\\
    \midrule
    GenericAds (Closed) & 3.92 & 2.24 & 9.46 & 3.98 & 9.51 & 0.9856 & 0.9452 & 0.4933 & 0.6549  \\
    NextAds (Closed) & 8.45 & 8.37 & 7.83 & 8.71 & 2.76 & 0.9681 & 0.9358 & 0.5377 & 0.6384  \\
    NextAds (Open) & 5.44 & 4.24 & 6.44 & 6.37 & 7.43 & 0.9533 & 0.9439 & 0.4440 & 0.5873  \\
    \rowcolor{gray!15} \textbf{MAPS} & \textbf{8.96} & \textbf{8.73} & \textbf{9.78} & \textbf{8.99} & \textbf{9.87} & \textbf{0.9886} & \textbf{0.9611} & \textbf{0.5504} & \textbf{0.6682}  \\
    \bottomrule
    \end{tabular}
    }
    \caption{Performance comparison across PCG-Bench and PCI-Bench.}
    \label{tab:overall_performance}
\end{table*}

\subsection{Director Agent}
The Director Agent is the creative realization module of the MAPS framework, responsible for transforming the personalized advertisement script produced by the Scripter Agent into a deliverable video creative. Its inputs come from two sources: the structured advertisement script output by the Scripter Agent, which fully specifies the script content, voiceover, visual descriptions, and pacing for each scene; and the available visual assets of the target product, including product images and brand logos, which serve as visual grounding constraints to ensure that the generated content faithfully reflects the product's visual characteristics and brand identity.

Upon receiving the above inputs, the Director Agent integrates the script content with the product visual assets to construct a complete multimodal input for the downstream video generation model. Specifically, scene descriptions, camera motion instructions, and voiceover text from the script are converted into textual prompts, while product images are simultaneously introduced as visual reference constraints, forming a multimodal conditioning input that guides the video generation model to synthesize personalized advertisement video content aligned with user preferences while maintaining product visual consistency.

\subsection{Reflector Agent}
The Reflector Agent serves as the feedback-driven refinement module of the MAPS framework, closing the loop between deployed creatives and upstream generation. It operates on online user interaction signals collected after each advertisement impression, such as play duration, likes, comments, and purchases. It attributes these signals to their underlying causes, converts the attribution results into minimal revision directives, and routes them back to the Profile Agent and the Scripter Agent for iterative refinement.

\subsubsection{Feedback Attribution.}
Given the online interaction signals associated with a deployed creative, the Reflector Agent performs attribution to localize the source of suboptimal performance, distinguishing between two complementary categories of causes:

\begin{itemize}
    \item \textbf{Profile-level causes}: inaccuracies in user modeling, such as misjudged topical interests, mismatched presentation styles, or incorrectly inferred decision-making tendencies, which indicate that the user profile fails to faithfully capture the user's actual preferences.

    \item \textbf{Script-level causes}: deficiencies in the generated script, such as a weak narrative hook, misaligned tone or pacing, unclear CTA, or suboptimal presentation of product selling points, which indicate that the script itself does not effectively engage the target user.
\end{itemize}

\subsubsection{Minimal Revision and Routing.}
Based on the attribution results, the Reflector Agent produces minimal revision directives that can be executed by upstream agents at low latency and bounded cost, avoiding full regeneration of the creative. Profile-level directives are routed back to the Profile Agent to update the structured user profile, whereas script-level directives are routed back to the Scripter Agent to iteratively refine the advertisement script. Through this targeted feedback loop, MAPS continuously adapts its creatives to evolving user preferences across successive impressions, enabling closed-loop creative evolution without re-running the entire generation pipeline from scratch.
\section{Experiments}
\label{sec:experiments}

\begin{figure*}[t]
    \centering
    \includegraphics[width=0.87\textwidth]{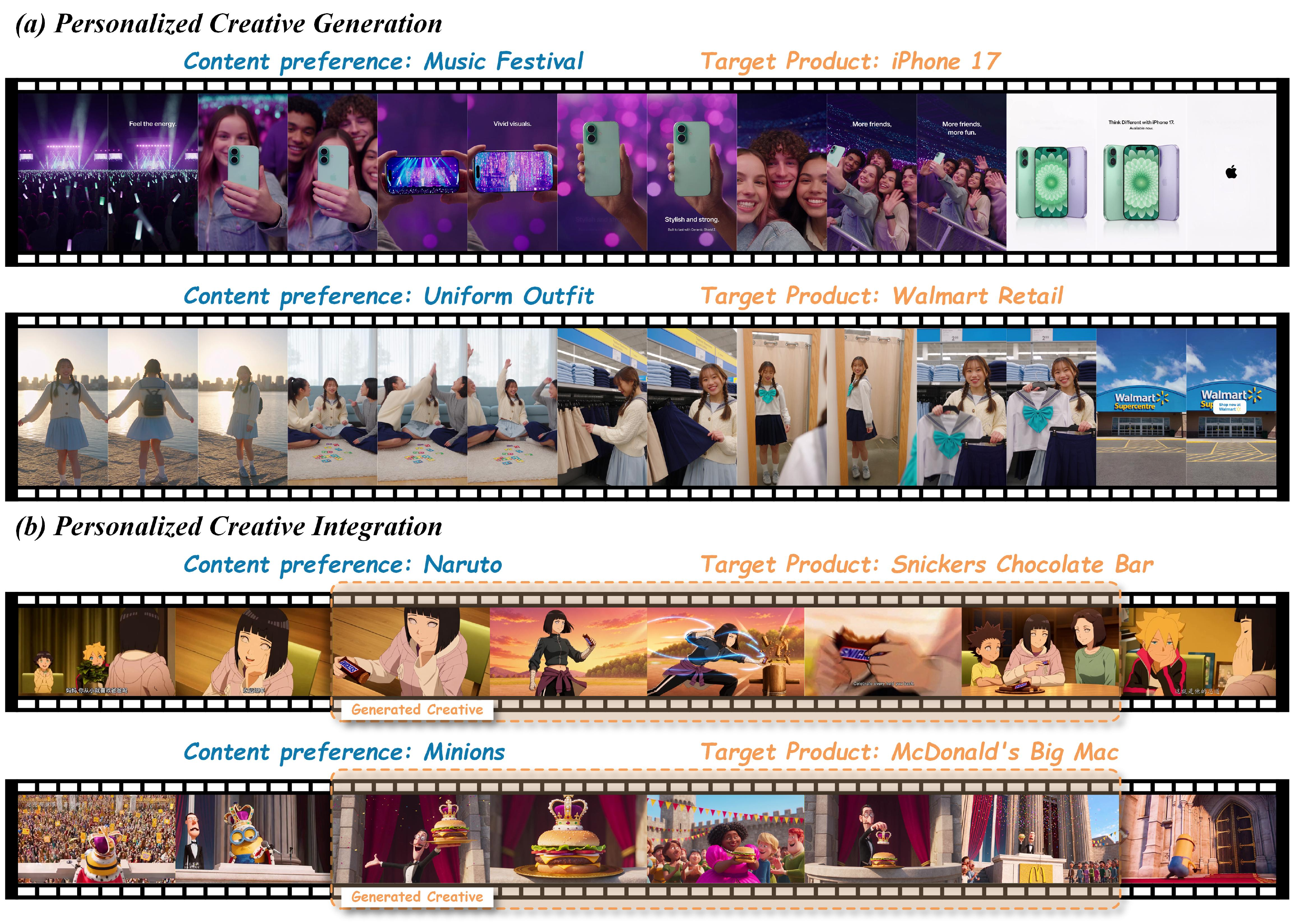}
    \caption{Demos of the PCG and PCI tasks.}
    \label{fig:PCG-PCI_demo}
\end{figure*}

\subsection{Datasets and Implementation Details}

\subsubsection{Datasets}
We conduct offline experiments on two representative benchmarks, PCG-Bench and PCI-Bench~\cite{xu2026nextads}, corresponding to the Personalized Creative Generation (PCG) and Personalized Creative Integration (PCI) tasks, respectively. The PCG task aims to synthesize standalone personalized video ad creatives, while the PCI task focuses on seamlessly embedding advertising messages into a host video currently being watched by the user in a non-intrusive manner. Both benchmarks include 50 users, each matched with 5 target products to form 250 user-product pairs, sharing a product library of 18 products. PCG-Bench is built upon the Qilin recommendation dataset~\cite{chen2025qilin}, while PCI-Bench is constructed from the MicroLens dataset~\cite{ni2025content}, with the last interacted video per user designated as the host video for creative integration.

\subsubsection{Implementation Details}

MAPS comprises four collaborative agents: the Profile Agent, built on Qwen3-VL-32B-Instruct~\cite{bai2025qwen3}, extracts multi-dimensional user preference profiles from historical interaction records; the Scripter Agent, with Qwen3-VL-8B-Instruct~\cite{bai2025qwen3} as its backbone, converts user profiles and product information into structured storyboard scripts; the Director Agent employs Kling-Omni~\cite{team2025kling} as the video generation model, taking the structured script as a conditional input to produce the final personalized ad video; and the Reflector Agent, built on Qwen3-VL-32B-Instruct~\cite{bai2025qwen3}, attributes online user feedback to profile-level and script-level causes and routes minimal revisions back to the Profile and Scripter Agents. The Scripter Agent is trained in two stages: the SFT cold start stage formulates the task as sequence-to-sequence generation to maximize the conditional log-likelihood of ground-truth scripts in a structured format, fine-tuning via LlamaFactory~\cite{zheng2024llamafactory} with LoRA~\cite{hu2022lora} of rank 32, using AdamW with a learning rate of $\text{1.0}\text{e}\text{-5}$, a batch size of 4, and a maximum sequence length of 8192 tokens for 20 epochs; the RL stage, implemented with the rLLM framework~\cite{tan2025rllm}, applies the GRPO algorithm to jointly align generated scripts with user preferences and product information, with a KL divergence coefficient of 0.04, AdamW optimizer at a learning rate of $\text{1.0}\text{e}\text{-6}$, a batch size of 4, for 5000 training steps. All experiments are conducted on 8 NVIDIA H100 GPUs.

\subsection{Offline Evaluation}

\subsubsection{Evaluation Metrics}
We design an evaluation suite covering five dimensions: video quality, personalization, product alignment, integration quality, and diversity. For video quality, we adopt Motion Smoothness (MS), Dynamic Degree (DD), Aesthetic Quality (AQ), and Imaging Quality (IQ) from VBench~\cite{huang2024vbench}. For the remaining dimensions, we employ GPT-5~\cite{singh2025openai} as a judge, scoring each criterion on a 0--10 scale with predefined rules. Specifically, personalization assesses the alignment between the generated creative and user historical preferences across presentation style and content preference; product alignment evaluates product identity accuracy, brand elements, and factual consistency; diversity measures variation among creatives generated for the same user across visual style and narrative structure; integration quality, applicable only to the PCI task, evaluates contextual flow and user experience disruption after inserting the creative into the host video. Detailed scoring prompts are provided in Sec.~\ref{appendix:eval} of the supplementary material.

\subsubsection{Comparison with State-of-the-Art Methods}

As shown in Table~\ref{tab:overall_performance}, MAPS achieves the best performance across all metrics on both benchmarks. On PCG-Bench, MAPS not only advances personalization and product alignment but also improves diversity from 6.69 to 7.30, and ranks first on all four video quality metrics. On the more challenging PCI-Bench, while NextAds~(Closed) improves personalization at the cost of product alignment (dropping from 9.46 to 7.83), MAPS mitigates this trade-off through structured profile extraction and script generation, simultaneously achieving the highest diversity score and the best integration quality. These results validate that multi-agent collaboration enables global optimization across competing objectives, mitigating the trade-offs commonly observed in existing baselines.

\subsubsection{Ablation Study}

\begin{table}[!t]
    \small
    \centering
    \resizebox{1.00\linewidth}{!}{
    \begin{tabular}{cccc|cc|c|c}
    \toprule
    \multicolumn{4}{c|}{\textbf{Agent Components}} & \multicolumn{2}{c|}{\textbf{Training}} & \multirow{2}{*}{\textbf{Pers.}} & \multirow{2}{*}{\textbf{Prod. Align.}}\\
    \textbf{Profile} & \textbf{Scripter} & \textbf{Director} & \textbf{Reflector} & \textbf{SFT} & \textbf{RL} & \\
    \midrule
     & & \ding{51} & & & & 3.10 & 7.83 \\
     & \ding{51} & \ding{51} & & & & 4.35 & 8.46 \\
     \ding{51} & \ding{51} & \ding{51} & & & & 6.92 & 8.67 \\
     \ding{51} & \ding{51} & \ding{51} & \ding{51} & & & 7.60 & 8.74 \\
     \ding{51} & \ding{51} & \ding{51} & \ding{51} & \ding{51} & & 7.81 & 9.12 \\
     \rowcolor{gray!15} \ding{51} & \ding{51} & \ding{51} & \ding{51} & \ding{51} & \ding{51} & \textbf{8.85} & \textbf{9.78} \\
    \bottomrule
    \end{tabular}
    }
    \caption{Ablation study on PCI-Bench.}
    \label{tab:ablation}
\end{table}

To validate the contribution of each component in MAPS, we conduct ablation studies on PCI-Bench, as shown in Table~\ref{tab:ablation}. Starting from a baseline with only the Director Agent, the incremental introduction of the Scripter Agent, Profile Agent, and Reflector Agent drives successive improvements in personalization, confirming their respective roles in enhancing informational consistency, enabling deep personalization, and refining creatives through feedback-driven attribution. Building upon the multi-agent collaboration, the SFT training stage establishes the structural competence of the Scripter Agent, laying a solid foundation for subsequent alignment. Finally, adding the RL stage, the full MAPS system achieves the best performance on both dimensions, improving personalization by 1.25 points and product alignment by 1.04 points compared to the variant without SFT and RL. This demonstrates that joint optimization via the multi-dimensional reward function drives the Scripter Agent to achieve synergistic alignment with both user preferences and product information, rather than trading one off against the other.

\subsection{Online Evaluation}


We further validate MAPS through online A/B testing in a specific scenario in the Kuaishou advertising system. We compare MAPS against two production baselines that have been successively deployed in the advertising pipeline. Deep Learning Recommendation Model (DLRM)~\cite{cheng2016wide,covington2016deep,zhou2018deep,zhu2018learning} is a retrieval-based recommendation model that matches users with creatives selected from a fixed content pool. 
Generative Recommendation Model (GRM)~\cite{deng2025onerec,xue2026generative} extends retrieval-based recommendation by modeling user interest as discrete semantic IDs (SIDs)~\cite{rajput2023recommender} through autoregressive generation, but uses the predicted SIDs as retrieval indices over a fixed content pool rather than generating new creatives.

We report three business-critical metrics: click-through rate (CTR), conversion rate (CVR), and average revenue per user (ARPU), where ARPU is defined as the total revenue divided by the number of users over the same period. As shown in Table~\ref{tab:online_evaluation}, MAPS improves CTR by 4.88\%, CVR by 4.01\%, and ARPU by 12.89\% over DLRM. We attribute this gain to the shift from retrieval to generation. DLRM can only match users with creatives drawn from a fixed content pool, and its static and limited inventory cannot keep pace with the evolving preferences of individual users. In contrast, MAPS generates personalized creatives on demand that fit each user and product more closely. MAPS further improves CTR by 2.87\%, CVR by 2.74\%, and ARPU by 9.85\% over GRM. Although GRM already leverages generative models to predict semantic IDs, it still retrieves creatives from a limited content pool. In contrast, MAPS produces native creatives on demand without being bounded by a pre-built inventory. The large ARPU improvement indicates that generating personalized creatives tailored to each user extends beyond better matching and unlocks commercial value that retrieval-based systems cannot reach with a limited creative pool, confirming the practical effectiveness of MAPS at industrial scale.

\begin{table}[!t]
    \small
    \centering
    \begin{tabular}{llll}
    \toprule
    \textbf{Method} & \textbf{CTR} & \textbf{CVR} & \textbf{ARPU} \\
    \midrule
    DLRM $\rightarrow$ MAPS & +4.88\% & +4.01\% & +12.89\% \\
    GRM $\rightarrow$ MAPS & +2.87\% & +2.74\% & +9.85\% \\
    \bottomrule
    \end{tabular}
    \caption{Online A/B evaluation results in a specific scenario in the Kuaishou advertising system.}
    \label{tab:online_evaluation}
\end{table}

\section{Conclusion}
\label{sec:conclusion}

In this paper, we introduce MAPS, a multi-agent personalized script-driven video advertising system for advertisement creative generation.
By coordinating the Profile, Scripter, Director, and Reflector Agents, MAPS transforms personalized video advertising from static retrieval over a limited creative inventory into dynamic generation within a continuous creative space. 
We apply reinforcement learning to script generation, enabling the production of structured scripts that integrate product information into narratives aligned with user preferences. 
Furthermore, MAPS supports continuous creative evolution across successive impressions by integrating post-deployment online user feedback into a closed-loop refinement mechanism. 
Extensive offline experiments on PCG-Bench and PCI-Bench demonstrate that MAPS outperforms existing methods across all evaluation metrics, and online A/B testing in a specific scenario in the Kuaishou advertising system further validates its effectiveness at industrial scale. Future work will explore finer-grained online feedback modeling and long-horizon creative evolution in larger-scale scenarios.

{
    \small
    \bibliographystyle{ieeenat_fullname}
    \bibliography{main}
}

\clearpage
\setcounter{page}{1}
\maketitlesupplementary

\section{Evaluation Prompts}
\label{appendix:eval}

To evaluate the generated creatives, we employ GPT-5~\cite{singh2025openai} as the LLM judge and design dedicated scoring prompts, as detailed below.

\begin{tcolorbox}[breakable, colback=gray!5!white, colframe=black!60!black, title=Judge Prompt for Personalization (Presentation)]
\textbf{You are an expert in evaluating the presentation style of advertisement creatives.}

Given a target product and the covers of the user's historically interacted videos, assess how well the generated advertisement aligns with the user's aesthetic preferences in visual presentation. Consider the following aspects:
\begin{itemize}[leftmargin=*]
    \item \textbf{Tone}: Does the overall tone match the user's preference (warm, cold, or neutral)?
    \item \textbf{Lighting}: Is the lighting treatment suited to the user's preference (soft, harsh, or natural)?
    \item \textbf{Texture}: Does the textural style (smooth, rough, \etc) fit the user's aesthetic inclination?
    \item \textbf{Artistic style}: Does the artistic style (minimalistic, modern, vintage, \etc) align with the user's taste?
\end{itemize}

\tcbline

\textbf{Scoring Criteria:}
\begin{itemize}[leftmargin=*]
    \item \textbf{Perfect match (10)}: The presentation style aligns perfectly with the user's preferences.
    \item \textbf{Good match (7--9)}: The presentation style aligns well, with only minor differences.
    \item \textbf{Moderate match (4--6)}: The presentation style partially matches the user's preferences.
    \item \textbf{Poor match (1--3)}: The presentation style does not align well with the user's preferences.
    \item \textbf{No match (0)}: The presentation style completely fails to align with the user's preferences.
\end{itemize}

\tcbline

\textbf{Output JSON:}
\begin{lstlisting}[style=jsonstyle]
{
  "score": number,
  "reason": "brief analysis"
}
\end{lstlisting}
\end{tcolorbox}

\begin{tcolorbox}[breakable, colback=gray!5!white, colframe=black!60!black, title=Judge Prompt for Personalization (Content)]
\textbf{You are an expert in evaluating how well an advertisement creative matches the user's content interests.}

Given a target product and the covers of the user's historically interacted videos, assess how well the generated advertisement aligns with the user's content preferences. Consider the following aspects:
\begin{itemize}[leftmargin=*]
    \item \textbf{Subject matter}: Does the subject matter of the advertisement align with the user's interest domains?
    \item \textbf{Scene semantics}: Does the scene selection (location, setting) fit what the user would enjoy?
    \item \textbf{Information angle}: Does the advertisement focus on the aspects the user cares about (performance, aesthetics, or value)?
    \item \textbf{Motivational appeal}: Does the advertisement appeal to motivations that resonate with the user (quality, efficiency, or lifestyle)?
\end{itemize}

\tcbline

\textbf{Scoring Criteria:}
\begin{itemize}[leftmargin=*]
    \item \textbf{Perfect match (10)}: The content aligns perfectly with the user's preferences.
    \item \textbf{Good match (7--9)}: The content aligns well, with only minor differences.
    \item \textbf{Moderate match (4--6)}: The content partially aligns with the user's preferences.
    \item \textbf{Poor match (1--3)}: The content does not align well with the user's preferences.
    \item \textbf{No match (0)}: The content is completely irrelevant to the user.
\end{itemize}

\tcbline

\textbf{Output JSON:}
\begin{lstlisting}[style=jsonstyle]
{
  "score": number,
  "reason": "brief analysis"
}
\end{lstlisting}
\end{tcolorbox}

\begin{tcolorbox}[breakable, colback=gray!5!white, colframe=black!60!black, title=Judge Prompt for Product Alignment]
\textbf{You are an expert in evaluating the consistency between an advertisement creative and the target product.}

Given the target product information and the generated advertisement, assess how faithfully the advertisement represents the product. Consider the following aspects:
\begin{itemize}[leftmargin=*]
    \item \textbf{Product identity accuracy}: Does the advertisement correctly portray the product's appearance, color, model, packaging, and accessories?
    \item \textbf{Brand and recognition elements}: Does the advertisement accurately present the product's logo, brand visual language, and design style?
    \item \textbf{Selling points and factual consistency}: Do the product claims conveyed in the advertisement reflect the real product features without exaggeration or misrepresentation?
\end{itemize}

\tcbline

\textbf{Scoring Criteria:}
\begin{itemize}[leftmargin=*]
    \item \textbf{Perfect match (10)}: The product in the advertisement matches the actual product in every respect.
    \item \textbf{Good match (7--9)}: The product is mostly consistent, with only minor details altered.
    \item \textbf{Moderate match (4--6)}: The product is partially consistent, but with some significant differences.
    \item \textbf{Poor match (1--3)}: The product in the advertisement is mostly inconsistent with the actual product.
    \item \textbf{No match (0)}: The product identity is completely inconsistent with the actual product.
\end{itemize}

\tcbline

\textbf{Output JSON:}
\begin{lstlisting}[style=jsonstyle]
{
  "score": number,
  "reason": "brief analysis"
}
\end{lstlisting}
\end{tcolorbox}

\begin{tcolorbox}[breakable, colback=gray!5!white, colframe=black!60!black, title=Judge Prompt for Integration Quality]
\textbf{You are an expert in evaluating how naturally an advertisement is integrated into the surrounding video content.}

Given a video in which the advertisement is embedded, assess how seamless or disruptive the advertisement feels to the viewing experience. Consider the following aspects:
\begin{itemize}[leftmargin=*]
    \item \textbf{Contextual flow}: Is the transition between the organic content and the advertisement smooth, or is it a sudden, jarring break?
    \item \textbf{Thematic consistency}: Does the advertisement share a similar tone, visual style, or topic with the surrounding content?
    \item \textbf{User experience disruption}: Is the advertisement placed at a logical breaking point, rather than interrupting a key moment?
    \item \textbf{Narrative integration}: Is the advertisement woven into the storyline, rather than appearing as an external overlay or hard cut?
\end{itemize}

\tcbline

\textbf{Scoring Criteria:}
\begin{itemize}[leftmargin=*]
    \item \textbf{Seamless (10)}: The advertisement is perfectly integrated and barely perceived as a disruption.
    \item \textbf{Natural (7--9)}: The advertisement is well placed and shares thematic elements with the video, causing minimal friction.
    \item \textbf{Noticeable but acceptable (4--6)}: The advertisement is clearly a break from the content, but occurs at a logical pause or maintains a similar aesthetic.
    \item \textbf{Intrusive (1--3)}: The advertisement feels forced and significantly breaks the immersion.
    \item \textbf{Highly disruptive (0)}: The advertisement completely ruins the viewing experience through poor timing or aggressive interruption.
\end{itemize}

\tcbline

\textbf{Output JSON:}
\begin{lstlisting}[style=jsonstyle]
{
  "score": number,
  "reason": "brief analysis"
}
\end{lstlisting}
\end{tcolorbox}

\begin{tcolorbox}[breakable, colback=gray!5!white, colframe=black!60!black, title=Judge Prompt for Diversity]
\textbf{You are an expert in evaluating the diversity of advertisement creatives.}

Given a set of advertisements generated for the same user across different products, assess how distinct these advertisements are from one another. Consider the following aspects:
\begin{itemize}[leftmargin=*]
    \item \textbf{Visual style}: Is there excessive similarity in tone, lighting, texture, and artistic style across the different product advertisements?
    \item \textbf{Narrative form}: Is the narrative structure (tutorial, vlog, plot twist, \etc) and tone (professional, humorous, \etc) overly similar across products?
    \item \textbf{Subject matter and scene semantics}: Is the subject matter or scene selection overly similar across the advertisements targeting the same user?
\end{itemize}

\tcbline

\textbf{Scoring Criteria:}
\begin{itemize}[leftmargin=*]
    \item \textbf{Fully diverse (10)}: The advertisements are completely distinct with no noticeable overlap.
    \item \textbf{Mostly diverse (7--9)}: The advertisements are mostly distinct, with only minor similarities.
    \item \textbf{Moderately diverse (4--6)}: The advertisements are somewhat similar but still distinct in key areas.
    \item \textbf{Slightly diverse (1--3)}: The advertisements are largely similar, with only slight variations.
    \item \textbf{Not diverse (0)}: The advertisements are identical in all aspects.
\end{itemize}

\tcbline

\textbf{Output JSON:}
\begin{lstlisting}[style=jsonstyle]
{
  "score": number,
  "reason": "brief analysis"
}
\end{lstlisting}
\end{tcolorbox}

\section{Qualitative Demos}
\label{appendix:demos}

\begin{figure*}[t]
    \centering
    \includegraphics[width=\textwidth]{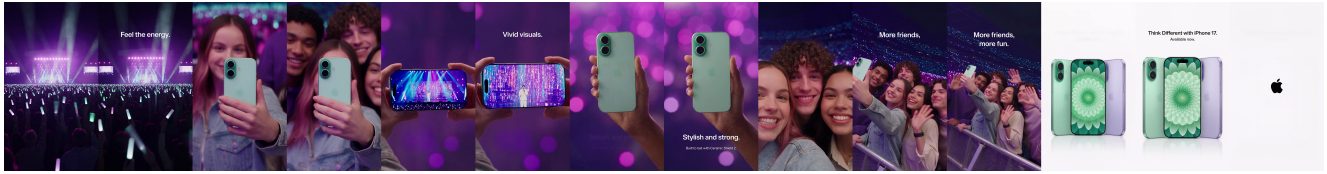}
    \caption{Demo of generated ad video. Content preference: Music festival. Target Product: iPhone 17.}
    \label{fig:supp_demo_1}
\end{figure*}

\begin{figure*}[t]
    \centering
    \includegraphics[width=\textwidth]{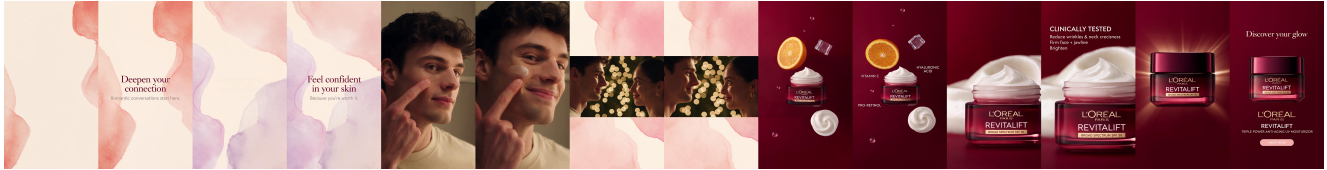}
    \caption{Demo of generated ad video. Content preference: Relationship topics. Target Product: L'Or\'eal Paris Revitalift.}
    \label{fig:supp_demo_2}
\end{figure*}

\begin{figure*}[t]
    \centering
    \includegraphics[width=\textwidth]{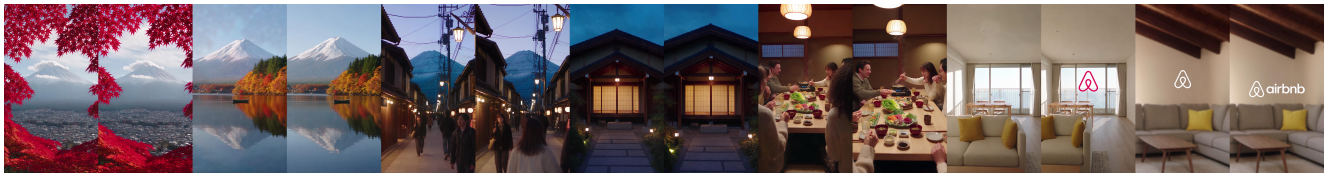}
    \caption{Demo of generated ad video. Content preference: Travel. Target Product: Airbnb.}
    \label{fig:supp_demo_3}
\end{figure*}

\begin{figure*}[t]
    \centering
    \includegraphics[width=\textwidth]{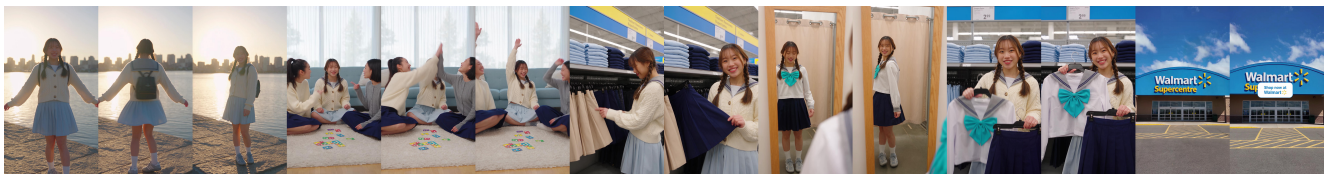}
    \caption{Demo of generated ad video. Content preference: Uniform outfit. Target Product: Walmart retail.}
    \label{fig:supp_demo_4}
\end{figure*}

\begin{figure*}[t]
    \centering
    \includegraphics[width=\textwidth]{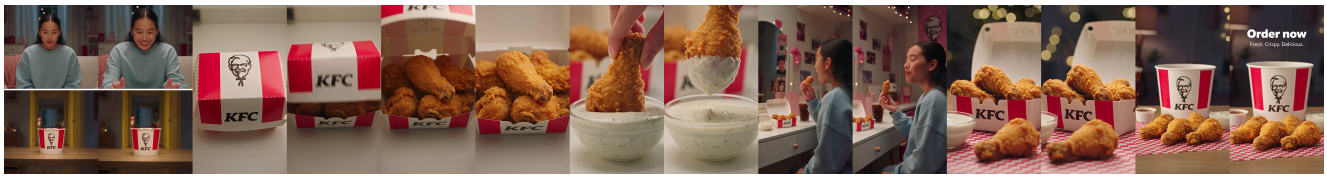}
    \caption{Demo of generated ad video. Content preference: Daily life. Target Product: KFC chicken nuggets.}
    \label{fig:supp_demo_5}
\end{figure*}

\begin{figure*}[t]
    \centering
    \includegraphics[width=\textwidth]{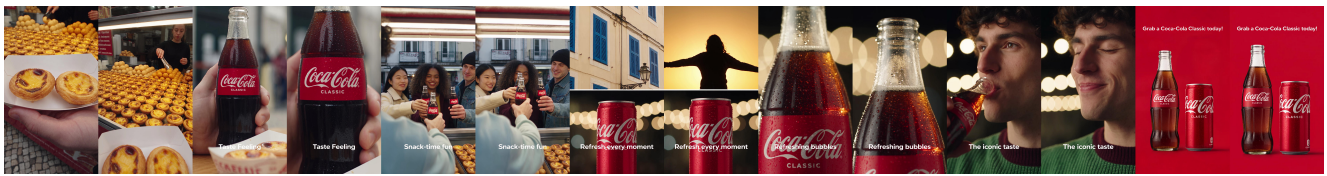}
    \caption{Demo of generated ad video. Content preference: Delicious food. Target Product: Coca-Cola.}
    \label{fig:supp_demo_7}
\end{figure*}

\begin{figure*}[t]
    \centering
    \includegraphics[width=\textwidth]{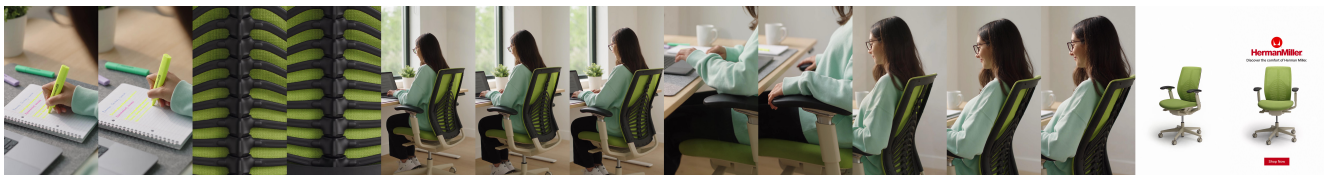}
    \caption{Demo of generated ad video. Content preference: Examination. Target Product: Herman Miller chair.}
    \label{fig:supp_demo_8}
\end{figure*}

We present qualitative demos of the ad videos generated by MAPS in Figures~\ref{fig:supp_demo_1}--\ref{fig:supp_demo_8}, each targeting a different user content preference and product.

\end{document}